\documentclass[conference]{IEEEtran}
\IEEEoverridecommandlockouts
\usepackage{cite}
\usepackage{amsmath,amssymb,amsfonts}
\usepackage{algorithmic}
\usepackage{graphicx}
\usepackage{textcomp}
\usepackage{xcolor}

\usepackage[english]{babel}
\usepackage[autostyle=true]{csquotes}
\usepackage[hyphens]{url}
\usepackage{hyperref}
\usepackage[capitalise,noabbrev]{cleveref}
\usepackage{comment}
\usepackage{siunitx}
\usepackage{mathtools}
\usepackage[ruled,norelsize]{algorithm2e}
\usepackage{standalone}
\usepackage{xspace}
\usepackage{booktabs,tabularx}
\usepackage{pgfplots}
\usetikzlibrary{patterns}

\makeatletter
\newcommand{\removelatexerror}{\let\@latex@error\@gobble}
\makeatother

\pgfplotsset{compat=newest,
    width=0.72\linewidth,
    height=0.35\linewidth,
}

\newcommand*{\Fregister}{\textsc{register()}\xspace}
\newcommand*{\Fdistshares}{\textsc{distribute\_shares()}\xspace}
\newcommand*{\Fdispute}{\textsc{submit\_dispute()}\xspace}
\newcommand*{\Fgenmpk}{\textsc{generate\_mpk()}\xspace}
\newcommand*{\Fsecret}{\textsc{submit\_secret()}\xspace}
\newcommand*{\Fsubmsk}{\textsc{submit\_msk()}\xspace}

\newcommand*{\bigprime}{p}

\newcommand*{\Group}{\mathbb{G}}
\newcommand*{\gen}{g}
\newcommand*{\Z}{\ensuremath{\mathbb{Z}_\bigprime}}
\newcommand*{\mpk}{\textit{mpk}\xspace}
\newcommand*{\msk}{\textit{msk}\xspace}
\newcommand*{\pk}{\textit{pk}\xspace}
\newcommand*{\sk}{\textit{sk}\xspace}

\newcommand*{\tmpk}{\tilde{\smash{\textit{mpk}}\rule{0pt}{1ex}}}

\newcolumntype{L}{>{\raggedright\arraybackslash}X}
\newcolumntype{R}{>{\raggedleft\arraybackslash}X}
\newcolumntype{C}{>{\centering\arraybackslash}X}

\begin{document}

\title{ETHTID: Deployable Threshold Information Disclosure on Ethereum\thanks{This work was supported by funding of the Helmholtz Association (HGF) through the Competence Center for Applied Security Technology (KASTEL).}}

\author{\IEEEauthorblockN{Oliver Stengele, Markus Raiber, Jörn Müller-Quade, Hannes Hartenstein}
\IEEEauthorblockA{\textit{Institute of Information Security and Dependability (KASTEL)} \\
\textit{Karlsruhe Institute of Technology}\\
Karlsruhe, Germany \\
\{oliver.stengele, markus.raiber, joern.mueller-quade, hannes.hartenstein\}@kit.edu}
}

\maketitle

\begin{abstract}
We address the Threshold Information Disclosure (TID) problem on Ethereum:
An arbitrary number of users commit to the scheduled disclosure of their individual messages recorded on the Ethereum blockchain if and only if all such messages are disclosed.
Before a disclosure, only the original sender of each message should know its contents.
To accomplish this, we task a small council with executing a distributed generation and threshold sharing of an asymmetric key pair.
The public key can be used to encrypt messages which only become readable once the threshold-shared decryption key is reconstructed at a predefined point in time and recorded on-chain.
With blockchains like Ethereum, it is possible to coordinate such procedures and attach economic stakes to the actions of participating individuals.
In this paper, we present ETHTID, an Ethereum smart contract application to coordinate Threshold Information Disclosure.
We base our implementation on ETHDKG~\cite{ethdkg}, a smart contract application for distributed key generation and threshold sharing, and adapt it to fit our differing use case as well as add functionality to oversee a scheduled reconstruction of the decryption key.
For our main cost saving optimisation, we show that the security of the underlying cryptographic scheme is maintained.
We evaluate how the execution costs depend on the size of the council and the threshold and show that the presented protocol is deployable on Ethereum with a council of more than 200 members with gas savings of 20--40\% compared to ETHDKG.
\end{abstract}

\section{Introduction}\label{sec:intro}
In this paper, we make use of threshold cryptography to achieve \emph{Threshold Information Disclosure} (TID) on Ethereum, an established public blockchain, when mutual trust cannot be assumed and protocol violations need to be discouraged.
Threshold Information Disclosure provides a way for arbitrarily many users to coordinate the disclosure of individually-held pieces of information:
Each user commits to the disclosure of their information if and only if the information of all other users is disclosed as well.
Before such a disclosure, each user should only be privy to their own piece of information.
Such a functionality is useful to allow users to generate and record their submissions independently from each other with the assurance that no user can prevent the disclosure of any particular submission.
Applications where independent submissions are useful include scientific evaluation of experimental raw data, security audits of software by independent groups, and sealed-bid auctions where the secrecy of losing bids is not required.
To accomplish such a coordinated information disclosure, an asymmetric key pair should be generated in such a way that a public encryption key becomes available immediately but the corresponding decryption key is threshold-shared among a council that is tasked with recovering and publishing said key at a predefined point in the future.
Submissions can then be encrypted and published in the present but only become collectively readable once the decryption key is published.
In the meantime, neither submitters nor individual members of the council coordinating the disclosure can decrypt any submissions.

The functionality described above is challenging to achieve on public blockchain systems like Ethereum due to a core aspect of their functionality:
It is vital that user-generated information spreads quickly through their global peer-to-peer networks in order to be recorded on their corresponding blockchains.
This very requirement makes it difficult for multiple parties to coordinate the release of information:
One party has to go first without knowledge of subsequent submissions and other parties can wait, observe the submissions of others, and potentially change their submission accordingly.
Individual \enquote{commit and reveal} schemes exist, e.g.\ based on a cryptographic hash function, to nullify any ordering advantages, but they introduce the possibility of a party refusing to reveal their commitment, which may be unacceptable depending on the use case.
In order to enable an arbitrarily large number of parties to commit to the coordinated public release of submissions to a blockchain, a delegation mechanism is therefore needed.
To preserve the notion of decentralisation inherent in public blockchain systems, such a mechanism can not rely on any trusted third party or centralised coordinator, either during setup or execution.
In order to be deployable, the entire mechanism must function within the constraints of the underlying blockchain system.

With ETHDKG, Schindler et al.~\cite{ethdkg,schindler2019distributed} demonstrated that a distributed key pair generation for a BLS signature scheme \cite{boneh2003aggregate} can be coordinated and recorded with an Ethereum smart contract.
The core of their construction is the well-known distributed key generation by Feldman \cite{feldmanPracticalSchemeNoninteractive1987}.
Our use case differs in two significant ways: First, the asymmetric key pair is \emph{only} intended for the encryption of user-defined information to keep it secret until disclosure; and second, we explicitly require a scheduled reconstruction of the threshold-shared secret key to facilitate this disclosure.
Based on our use case and the corresponding cryptographic schemes, we can adapt ETHDKG to save on deployment and execution costs while extending it with the functionality necessary to orchestrate the scheduled key reconstruction.
In particular, we show that biasing attacks as described by Gennaro et al.~\cite{gennaro1999secure,gennaro2003secure}, against which ETHDKG employs a countermeasure by Neji et al.~\cite{neji2016distributed}, are of no concern in our case and we can therefore simplify the overall protocol to save costs.

The main contributions of this paper are as follows:
\begin{itemize}
    \item An analysis of the problem of coordinating a distributed key generation, threshold sharing, and scheduled reconstruction in a context of mutual distrust via a decentralised coordinator that can enforce economic punishments to achieve Threshold Information Disclosure.
    We derive requirements independent of the underlying cryptographic protocol or execution platform.
    \item A use-case oriented examination of cryptographic schemes with the goal of minimising deployment and execution costs of our implementation.
    \item An argument that our main optimisation maintains the security of the underlying cryptographic scheme.
    \item ETHTID, an Ethereum smart contract implementation to achieve Threshold Information Disclosure.
    \item An evaluation of ETHTID with regard to deployment and execution costs to determine its feasibility and limits.
    In this way, we demonstrate that a coordinated information disclosure mechanism is deployable on Ethereum, paving the way for new decentralised smart contract applications that were previously impossible.
\end{itemize}

The remainder of this paper is structured as follows:
We derive general requirements for a decentralised TID coordinator and give a system overview in \cref{sec:psso} before we address related work in \cref{sec:rw}.
In \cref{sec:bb}, we review the essential building blocks we make use of and give a brief rationale for their selection before we describe our implementation in \cref{sec:ethtid}.
We thoroughly evaluate our implementation in \cref{sec:eval} and discuss the results and open issues in \cref{sec:disc}.
With \cref{sec:concl}, we conclude.

\section{Problem Statement and System Overview}\label{sec:psso}
It is instructive to first consider the problem at hand separate from an implementation.
In this section, we derive requirements for the coordination activities and underlying cryptographic primitives with respect to  distributed key generation, threshold sharing, and reconstruction with economic punishments for the purpose of TID. In addition, we present a system overview. We assume the reader is familiar with the basic concepts of threshold cryptography.

Threshold Information Disclosure is based on
a \emph{threshold council} of \(n\) members tasked with generating a master public key \(\mpk\) such that the corresponding secret key \(\msk\) can be recovered after a predefined point in time \(r\) by the cooperation of at least \(t+1\) council members. TID must satisfy the following safety and liveness conditions: The master secret key \(\msk\) is not available to any party before time \(r\) (safety) and the master secret key \(\msk\) eventually becomes available to all  parties (liveness).

Note that both conditions refer to parties both actively participating in the generation, sharing, and reconstruction phases of threshold cryptography (i.e.\ the council members) as well as user parties that may just use \(\mpk\) and \(\msk\).
However, whether or not the safety and liveness conditions hold depends entirely on the behaviour of the council members.
Thus, the trust assumption of TID is that the council does not reconstruct the master secret key \(\msk\) before time \(r\).
The prominent means for achieving such behaviour in practice is an incentive scheme like the one presented by Yakira et al.~\cite{yakira2019rational}, coupled with the assumption that council members act rationally.
While a properly designed incentive scheme is use case dependent and, therefore, out of scope of this paper, the necessary functionality to put such an incentive scheme into practice is made feasible by the proposed smart contract ETHTID.

The threshold cryptographic concept works as follows.
First, a distributed key generation is performed where every council member contributes randomness to the public key.
This leads to an \((n,n)\)-sharing of the key, i.e.\ every piece is necessary for recovery.
Second, the members exchange information to obtain a \((t+1,n)\)-sharing of the key, so that only \(t+1\) out of \(n\) pieces are required for the scheduled recovery.
Since every member acts as the dealer of a \((t+1,n)\) secret sharing scheme in this step and none of them trust each other, we require verifiable secret sharing (VSS)~\cite{feldmanPracticalSchemeNoninteractive1987,pedersenNonInteractiveInformationTheoreticSecure1992}.
VSS prevents malicious members from distributing shares in such a way that some subsets of \(t+1\) members recover an incorrect secret key.
The last phase is the coordinated recovery of \(\msk\).

It is worth noting that the primary reason for performing a threshold sharing in the first place is to eliminate single points of failure and attain a certain degree of resilience.
Without it, if even one council member refuses to cooperate in the recovery, the master secret key is lost.
Once the threshold sharing is performed, \(n-t-1\) members can fail, be bribed or coerced and \(\msk\) can still be recovered.

As outlined above, council members have to broadcast and store certain pieces of information to hold each other accountable.
In case a member misbehaves, evidence is broadcast to the group and each member can act accordingly after verifying for themselves whether or not the accused member actually misbehaved.
This functionality now needs to be covered by a (decentralised) coordinator which, in addition to performing any necessary verification, should also hold and distribute deposits and rewards accordingly.
Therefore, a coordinator has to fulfil the following requirements:
\begin{enumerate}
    \item The coordinator must be able to store, and make available data from council members, thus allowing them to execute the VSS protocol and detect misbehaviour.
    \item The coordinator must be able to perform verification given previously stored data and submitted evidence.
    \item The coordinator must be able to receive, hold, and redistribute assets according to the outcome of verification.
    \item The coordinator must have access to a public form of timekeeping to enforce a schedule.
\end{enumerate}
While the first requirement can be fulfilled by any form of \enquote{bulletin board}-style public storage system, the last three requirements lead to the consideration of blockchain-based consensus systems like Ethereum.
In addition to their ability to perform verification and distribute assets accordingly, these systems are decentralised and include a timekeeping mechanism in the form of block height.
A smart contract can therefore act as a coordinator without becoming a single point of failure.
When selecting cryptographic primitives and a platform for the coordinator, the amount of information that must be stored and the difficulty of performing verification become crucial.
In \cref{sec:bb}, we present our selection of components and explain our design decisions.

On the side of the council members, we merely assume that they have access to the coordinator as a broadcast channel and are capable of securely storing certain pieces of data until the time for disclosure.
Note that we cannot prevent any parties from communicating privately with each other without involving the coordinator.
In practice, there should be mechanisms in place to discourage collusion, like the construction by Dong et al. \cite{dong2017betrayal}.

\begin{figure}[!t]
    \centering
    \includegraphics[width=0.95\columnwidth]{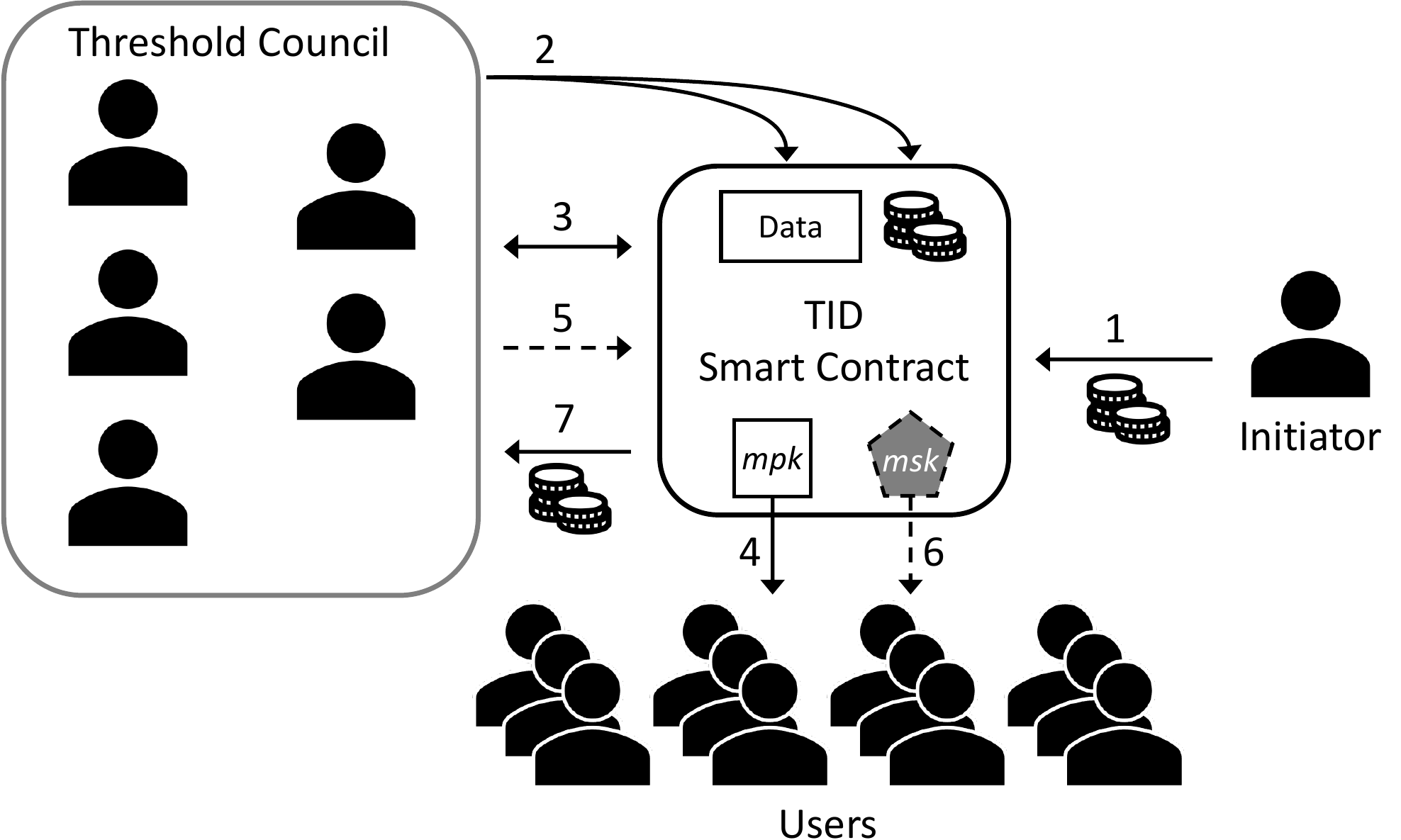}
    \caption[General system overview]{General system overview.
    (1) Initiator deploys TID smart contract with parameters and incentives.
    (2) Threshold council members register with the contract by submitting data and a deposit.
    (3) Council members communicate through the smart contract with each other to generate the public encryption key \(\mpk\) and establish a threshold sharing of the secret decryption  key \(\msk\). In case of misbehaviour, members submit a dispute to the TID contract to enforce punishments.
    (4) Users can obtain \(\mpk\).
    (5) At a codified time, council members submit data to the contract to enable the reconstruction of \(\msk\).
    (6) Users and the general public can obtain the decryption key \(\msk\).
    (7) Based on their behaviour, council members receive a reward in addition to their collateral and are released from their obligation.}
    \label{fig:overview}
\end{figure}
With the general structure and requirements in mind, we can now outline a system overview as well as the corresponding workflow.
\cref{fig:overview} shows the parties and their interactions.

\textit{Setup and registration phase:} An  \emph{initiator} triggers and finances the entire procedure.
While instantiating the TID contract, the initiator sets the threshold as a fraction of registered council members and provides a financial incentive for participation.
Recall from \cref{sec:intro} that Threshold Information Disclosure is a a time-sensitive service.
As such, the contract also enforces a schedule, not just for the reconstruction, but for all preceding phases as well.
In addition, the council members have to register with the smart contract.
The primary task of the threshold council is to collectively \enquote{keep a secret} since a smart contract is unable to perform this function itself.

\textit{Share distribution and dispute phase:} Once authorised, the council members perform a distributed key generation and submit data to the contract to hold each other accountable.
If they behave correctly, their submitted data serves to prove their honesty and protects them against false accusations.
However, if they misbehave, the same data, in conjunction with submitted evidence, will convict them and lead to economic punishment.
Alongside this verification data, council members may also need to send a security deposit to the contract that can be destroyed or redistributed in case they misbehave.
Next, the threshold sharing of the key is performed.
Once the council members are registered with the contract, they can privately exchange data through the smart contract to establish a \((t+1,n)\)-sharing of the previously generated key.
There are two cases of misbehaviour that can occur during this phase:
Either a member sends incorrect information or they do not send any information even though they are supposed to.
While these two cases are handled explicitly in protocols like the one by Pedersen~\cite{pedersenThresholdCryptosystemTrusted1991}, in practice they reduce to just the second case:
Rather than sending incorrect information, knowing that the recipient will immediately realise this and trigger punishments, a mischievous member will rather not send anything.
They can also be handled in a similar way by the coordinator:
A council member that registered but failed to broadcast the necessary data in time loses their deposit and is disqualified from the remaining protocol.
Likewise, a member that broadcast incorrect information via the coordinator will lose their deposit and be disqualified once another member files a valid dispute.
After the threshold sharing is established and any misbehaving members have been disqualified via a dispute, the data stored in the contract can be used to derive the public encryption key \(\mpk\), which the \emph{users} can then obtain to encrypt their messages.

\textit{Reconstruction phase:} Once the predefined time for reconstruction \(r\) has come, members of the council are incentivised to submit data necessary for the reconstruction of \(\msk\) to the contract.
If at least \(t+1\) members cooperate, the secret key can be recovered via the previously established threshold sharing.
A member can then post \(\msk\) to the coordinator and both the users and the general public can obtain it to decrypt all previously published messages, thus achieving a coordinated disclosure.
Lastly, the members of the threshold council can reclaim their deposit in addition to their share of the reward for their service.

\section{Related Work}\label{sec:rw}
Coordinating the release of information has previously been tackled in two distinct ways, either through time-lock puzzles or with trusted third party custodians in either centralised or decentralised fashion.
When comparing the work presented here with previous approaches, it is helpful to consider the number of parties that can coordinate the timed release of information without disclosing it to each other and to how many other parties the information is disclosed to.

Time-lock puzzles~\cite{RSW96,maoTimedReleaseCryptography2001,mahmoodyTimeLockPuzzlesRandom2011} bear similarities to our use case in that they can \enquote{send information into the future}.
The receiver of a puzzle is required to do some inefficient but feasible computation that is expected to take at least a certain amount of time to solve it.
The time it takes a recipient to solve the puzzle and recover the decryption key can only be estimated by the sender.
Additionally, the recipient must exert computational effort to eventually receive the encrypted information.
These schemes generally constitute a \enquote{one-sender-to-one-recipient} timed release functionality with low timing accuracy but without a trusted third party.

Another solution to timed-release encryption makes use of a trusted party that releases decryption keys at the right time \cite{RSW96,bellareVerifiablePartialKey1997,dicrescenzoConditionalObliviousTransfer1999,cathaloEfficientNoninteractiveTimedRelease2005}.
To circumvent a single point of failure that a trusted party poses, it is possible to securely distribute the task to multiple parties, such that it is sufficient that some amount of them behave honestly.
Some of these works allow for multiple recipients with more or less favourable scalability or setup procedures, but lacking a distributed key setup they only support a timed release for individual senders.

Benhamouda et al.~\cite{benhamoudaCanBlockchainKeep2020} present a concept that allows the Algorand blockchain itself to act as a long-term secret keeper through a randomised sequence of anonymous committees.
They list functionalities like the one we present here as future achievable goals of their concept.
While their approach is more general and possibly covers the TID functionality, our approach is practically feasible on an established blockchain.
However, their \enquote{cryptographic sortition} approach to forming committees may be transferable to our concept as a countermeasure to Sybil attacks.
It will be interesting to see the future developments of Algorand and its capabilities.

In 2018, two projects appeared independently of each other within the Ethereum ecosystem that combined threshold secret sharing with Ethereum smart contracts: Kill Cord\footnote{\url{https://github.com/nomasters/killcord}} and Kimono\footnote{\url{https://medium.com/@pfh/kimono-trustless-secret-sharing-using-time-locks-on-ethereum-8e7e696494d}}. While the development of Kill Cord seems to be ongoing, Kimino appears to be abandoned and defunct.
Both use Shamir's secret sharing~\cite{shamirHowShareSecret1979} to fragment a decryption key and entrust the resulting shares to Ethereum nodes which are then incentivised via smart contracts to only post their shares at a certain time.
In the case of Kill Cord, the creator of such a key can delay the designated time of disclosure by \enquote{checking in} to a smart contract, thus providing the functionality of a deadman switch.
However, in both cases, the initiator of the protocol generates the key and acts as a trusted dealer, making both systems \enquote{one-to-many} compared to our construction which achieves a coordinated \enquote{many-to-many} release.

\section{Background: Building Blocks}\label{sec:bb}
When implementing the coordinator as a smart contract on a public blockchain, it is important to minimise the incurred operational costs.
These are (1) broadcast information which may be stored on the blockchain entirely or partially and (2) computations performed by the smart contract, mostly during optional verification.
Thus, the following building blocks were chosen according to these two metrics while still fulfilling the requirements mentioned in \cref{sec:psso}.

First, we need a public key encryption scheme.
The obvious candidates are the ElGamal and RSA cryptosystems.
We chose ElGamal for several reasons:
It can be used over elliptic curves thus achieving short public and secret keys;
more importantly it is easy to generate a distributed public key for it.
Generation of RSA keys involves probabilistic prime number generation and typically requires several iterations until suitable primes are found.
While there are several works on distributed key generation for RSA~\cite{bonehEfficientGenerationShared1997,bonehEfficientGenerationShared2001,hazayEfficientRSAKey2012,hazayEfficientRSAKey2019,frederiksenFastDistributedRSA2018}, they are much less efficient than the distributed key generation protocol for ElGamal due to Feldman~\cite{feldmanPracticalSchemeNoninteractive1987}, based on the threshold secret sharing scheme by Shamir~\cite{shamirHowShareSecret1979}.
It is important to note that, while we actually employ this primitive on an elliptic curve, we will use the more common notation of multiplicative groups here.
In the second part of this section, we briefly cover the aspects of Ethereum that are necessary for our implementation and evaluation.

\subsection{Distributed Key Generation}\label{sec:dkg}
Our goal is to generate an ElGamal public key for which the corresponding secret key is \((t+1,n)\)-secret shared among the council members that generated it.
We first briefly recall Shamir's threshold secret sharing scheme~\cite{shamirHowShareSecret1979}:
To share a secret \(s\), a random polynomial \(p\) of degree \(t\) with \(p(0) = s\) is drawn and each party \(i\) is given \(p(i)\) as its share.
Since \(p\) has degree \(t\), it is possible to uniquely reconstruct \(p\), and thus find \(s\), given \(t+1\) such shares, but \(t\) shares reveal no information at all about \(s\).
Handling data related to these polynomials is at the very core of the TID coordinator.

To generate a threshold shared ElGamal key, we use a variant of the Joint-Feldman DKG~\cite{feldmanPracticalSchemeNoninteractive1987} similar to the one by Pedersen~\cite{pedersenThresholdCryptosystemTrusted1991}, which we also briefly recall.
One important difference between these VSS protocols in theory and our practical implementation is the use of private communication channels between members.
In order for the TID coordinator to efficiently arbitrate disputes, all communication between members must flow through it.
We therefore adopt from ETHDKG \cite{ethdkg} a symmetrical encryption scheme for the confidential exchange of information between members and a non-interactive zero-knowledge proof and verification scheme as a part of dispute resolution.
The end result is that a dispute can be completed non-interactively in a single transaction.

We can now review the cryptographic procedures at the core of our implementation.
Let \(\Group\) be a cyclic group of prime order \(\bigprime\) with generator \(\gen\) for which the decisional Diffie-Hellman problem is hard.
First note that it is easy to generate an ElGamal public key where each member holds an additive share of the secret key and all \(n\) council members have to cooperate to reconstruct it:
Each party \(i\) chooses \(\sk_i \in \Z\), calculates \(\pk_i \coloneqq \gen^{\sk_i}\) and broadcasts \(\pk_i\).
The master public key is then \(\mpk \coloneqq \prod_{i=1}^n \pk_i = \prod_{i=1}^n \gen^{\sk_i} = \gen^{\sum_{i=1}^n \sk_i}\) and the secret key \(\msk \coloneqq \sum_{i=1}^n \sk_i\) is \((n,n)\)-secret shared among all members.
This is quite beneficial in our use case, because it allows executing the distributed key generation and threshold sharing simultaneously.
With this construction, it is also possible to disqualify members for misbehaviour and simply discard their pieces of \(\msk\) and \(\mpk\) as outlined below.

To achieve a \((t+1,n)\)-threshold secret sharing of the (not yet reconstructed) master secret key \(\msk\), each member individually \((t+1,n)\)-shares its contribution \(\sk_i\) among all \(n\) council members using Shamir secret sharing.
We will use the term \emph{shadow} for shares of \(\sk_i\), while reserving the term \emph{share} for shares of the master secret key \(\msk\).
Since Shamir secret sharing is additively homomorphic, i.e.\ adding up shares of two secrets and then performing a reconstruction yields the same result as reconstructing the secrets individually and then adding them together, each member can combine all valid shadows that they received into a single share of the master secret key.
To prevent malicious members from sending inconsistent shadows during this phase, we employ the following consistency checks:
When members pick a random polynomial \(p_i = \sk_i + \sum_{k=1}^{t} a_{i,k}X^k\), they also compute verification values \(A_{i,k} = \gen^{a_{i,k}}\) and broadcast \(\{A_{i,k}\}_{k=1}^{t}\) in addition to \(\pk_i\).
They also privately send to member \(j\) the shadow \(u_{i,j} = p_i(j)\) and treat \(p_i(i)\) as their own shadow.
When member \(j\) receives a shadow \(u_{i,j}\) from member \(i\), they check whether it agrees with the polynomial broadcast:
\begin{equation}\label{eq:shadow}
    \gen^{u_{i,j}} \overset{?}{=} \pk_i \prod_{k=1}^{t} A_{i,k}^{j^k} = \gen^{sk_i + \sum_{k=1}^{t}a_{i,k}j^k} = \gen^{p_i(j)}
\end{equation}
If this check fails, the party reveals the invalid shadow \(u_{i,j}\) to the coordinator and party \(i\) is excluded if the complaint is valid, i.e.\ the shadow is inconsistent with the corresponding polynomial broadcast.
Notice that the coordinator needs access to the verification values \(\pk_i\) and \(\{A_{i,k}\}_{k=1}^{t}\) of the accused member \(i\) in order to evaluate \cref{eq:shadow} as part of a dispute.
Rather than storing these values on-chain, we adopt the strategy of ETHDKG to store a hash of the previous broadcast and let the accusing member resubmit it.
Checking \cref{eq:shadow} ensures that all shadows \(u_{i,j}\) lie on the same degree \(t\) polynomial and thus reconstruction from any \(t+1\) shadows will yield the same value, namely \(\sk_i\).

Finally, each party \(i\) adds up all \(n\) shadows \(u_{i,j}\) it has, where \(n-1\) shadows were received from other parties and one comes from evaluating its own polynomial, to get a share \(s_i\) of the polynomial \(p = \sum_{i=1}^n p_i\), for which it holds that \(p(0) = \sum_{i=1}^n p_i(0) = \sum_{i=1}^n \sk_i = \msk\).
Since polynomial \(p\) is of degree \(t\), \(\msk\) can be recovered by a group of \(t+1\) cooperating members \({i_1,\ldots,i_{t+1}}\) through Lagrange interpolation:
\begin{equation}\label{eq:lagrange}
    \msk = \sum_{k=1}^{t+1} \left( \prod_{l \neq k} \frac{i_l}{i_l-i_k} s_{i_k} \right).
\end{equation}
With any fewer shares, no useful information of \(\msk\) can be derived.
Fortunately, \cref{eq:lagrange} does not need to be evaluated by the coordinator.
Instead, a council member can submit \(\msk\) and the coordinator can verify it against the master public key \(\mpk\).

To transmit shadows confidentially over the coordinator, Schindler et al.~\cite{ethdkg} employ a symmetric key encryption scheme inspired by the Diffie-Hellman key exchange \cite{rescorla1999rfc2631} and ElGamal encryption~\cite{elgamalPublicKeyCryptosystem1985}.
At the beginning of the protocol, every council member \(i\) submits a public encryption key \(\hat{\pk}_i = \gen^{\hat{\sk}_i}\) to the coordinator while keeping \(\hat{\sk}_i\) to themselves.
Note that \(\hat{\sk}_i\) is distinct from \(\sk_i\), the contribution of member \(i\) to the master secret key.
Every council member \(i\) can then compute a symmetric encryption key for another council member \(j\) as:
\begin{equation}\label{eq:symkey}
    k_{ij} = \hat{\pk}_j^{\hat{\sk}_i} = \hat{\pk}_i^{\hat{\sk}_j} = \gen^{\hat{\sk}_i \hat{\sk}_j}.
\end{equation}
With this key, shadows can then be encrypted before being broadcast via the coordinator.

In order to file a dispute, a council member \(j\) reveals to the coordinator the symmetric encryption key \(k_{ij}\) used by member \(i\) whom he accuses of sending an incorrect shadow.
To prove the correctness of this key without revealing \(\hat{\sk}_j\), Schindler et al.~\cite{ethdkg} employ a non-interactive zero knowledge proof and verification scheme that we adopt without any changes.
For the purpose of this paper, it suffices to know that unfounded disputes with an incorrect decryption key are not accepted by the coordinator.
By resubmitting the broadcast message of the accused member as part of the dispute, the coordinator can decrypt the shadow in question, check its validity against the verification values of the sender \(i\) with \cref{eq:shadow}, and disqualify member \(i\) if the shadow is found invalid.

\subsection{Ethereum}
With the previous section covering the cryptographic procedure we employ, we now turn to the environment and mechanisms to realise it.

We assume general familiarity with Ethereum~\cite{Buterin:2014wm} including transactions, the gas cost mechanism, and the underlying blockchain and peer-to-peer infrastructure.
The components of Ethereum that are most crucial for our implementation are \emph{precompiled contracts} for the elliptic curve operations that we rely on.
The term is rather misleading since they are not actually smart contracts that were compiled and stored at specific addresses on the blockchain, but rather optimised implementations of certain functions external to the EVM that are called as if they were contracts.
In our case, we rely on two precompiled contracts for addition and scalar multiplication on the Barreto-Naehrig~\cite{BNCurve} curve as defined by EIP-196\footnote{\url{https://eips.ethereum.org/EIPS/eip-196}}, which were deployed with the Byzantium hardfork in 2017.
At the end of 2019, the gas costs for both of these contracts were substantially reduced per EIP-1108\footnote{\url{https://eips.ethereum.org/EIPS/eip-1108}}, which was deployed as part of the Istanbul hardfork.

At the time of writing, the amount of gas that can be spent within a single block is approximately \num{15000000}, any transaction that exceeds this limit is practically impossible to execute.
This gas limit per block can be influenced by miners and therefore changes over time.
Lastly, the \emph{gas price} is a mechanism to incentivise miners to include transactions into blocks.
While the exact amount of gas that a transaction consumes is determined through execution, the gas price gives miners an easy way to prioritise transactions.
The higher the gas price, the more a miner is rewarded for including it into their block, regardless of how much gas the transaction actually consumes during execution.

\section{ETHTID}\label{sec:ethtid}
\begin{table}[!t]
    \centering
    \caption{Notation Overview.}
    \begin{tabularx}{0.95\linewidth}{lL}
    \toprule
    Symbol & Description\\
    \midrule
    \(n\) & Size of threshold council. \\
    \(t\) & Threshold. \(t+1\) cooperating council members can reconstruct the shared secret. \\
    \(r\) & Time for reconstruction of the shared secret. \\
    \(\langle \mpk, \msk \rangle\) & Master public and secret key. The latter is threshold-shared among the council. \\
    \(\langle \pk_i, \sk_i \rangle\) & Contributions of council member \(i\) to \(\mpk\) and \(\msk\).\\
    \(u_{ij}, \overline{u_{ij}}\) & Decrypted and encrypted shadow from council member \(i\) to member \(j\). \\
    \(s_i\) & Share of \(\msk\) held by council member \(i\).\\
    \(\langle \hat{\pk}_i, \hat{\sk}_i \rangle\) & Key pair of council member \(i\) used to generate symmetric key \(k_{ij}\). \\
    \(k_{ij}\) & Symmetric key between council members \(i\) and \(j\), used to encrypt shadows. \\
    \(\pi(k_{ij})\) & Zero knowledge proof of correctness for \(k_{ij}\), submitted as part of dispute. \\
    \bottomrule
    \end{tabularx}
    \label{tab:notation}
\end{table}
In this section, we describe ETHTID, a deployable concept for a decentralised Threshold Information Disclosure procedure as an Ethereum smart contract, based on ETHDKG~\cite{ethdkg,schindler2019distributed}.
We inherit the general structure from ETHDKG but optimise (simplify) and extend the implementation for our use case and requirements.
To facilitate reading as well as the mapping between \cref{sec:bb} and \cref{sec:ethtid}, we provide an overview of our notation in \cref{tab:notation}.

The goal of ETHDKG was to establish a group BLS signature \cite{boneh2003aggregate}, whereas we only need an ElGamal key pair for encryption and eventual decryption.
The first optimisation is rather straightforward: We can do without the bilinearity of the BN curve.
The main benefit is that the ETHTID contract does not need to perform any pairing checks, thus saving costs.
The reason we still use the BN curve is that it is currently the only elliptic curve available on Ethereum.
Using a simpler curve by implementing the necessary operations ourselves would be prohibitively expensive.

The second optimisation is more intricate but also more significant in terms of cost savings.
Gennaro et al. \cite{gennaro1999secure,gennaro2003secure} thoroughly examined how an attacker can bias the result of a distributed key generation by entering under the guise of multiple identities and then selectively denouncing some of them.
The core issue is that all information necessary to compute the result of the protocol, \(\mpk\) in our case, is known before the last opportunity to disqualify participants.
In ETHDKG \cite{ethdkg}, Schindler et al. employed a countermeasure by Neji et al. \cite{neji2016distributed} that involved an additional round of broadcasts and an optional reconstruction in case any party did not perform this broadcast.
Instead of preventing biasing attacks in this way, we can simply accept them.
This is because they do not impact the security of the ElGamal encryption scheme:
Assume an attacker waits and observes all \(\pk_i\) from honest parties, computing \(\tmpk = \prod \pk_i\).
The attacker may then choose any value \(b\) and force the resulting public key of the protocol to be \(\mpk = \tmpk \cdot g^b\).
Whatever attack an adversary can perform against this biased public key \(\mpk\) can also be performed against the unbiased public key \(\tmpk\).
Given \(\tmpk\) and a cipher text \(\tilde{c} = (\tilde{c}_1, \tilde{c}_2) \coloneqq (g^r, \tmpk^r \cdot m)\), and given \(b\), this cipher text can be transformed to the biased public key \(\mpk\): \(c = (c_1, c_2) \coloneqq (\tilde{c}_1, \tilde{c}_2 \cdot \tilde{c}_1^b) = (g^r, \tmpk^r \cdot g^{rb} \cdot m) = (g^r, (\tmpk \cdot g^b)^r \cdot m)\) which is a valid cipher text under \(\mpk = \tmpk \cdot g^b\).
Simply put, an attacker gains no advantage from biasing the ElGamal key pair, so we can simplify our protocol to save costs.
It is crucial to note that this optimisation is only viable because the key pair in question is meant to be used solely for ElGamal encryption.

\begin{table*}[!t]
\caption{Interface of the ETHTID Contract.}
\label{tab:tidinterface}
\renewcommand{\arraystretch}{1.3}
\begin{tabularx}{\textwidth}{p{0.55\columnwidth}X}
\toprule
Function & Description \\
\midrule
\textsc{register}(\(\hat{\pk}_i\)) & Council member commit to the participation in the protocol and submit \(\hat{\pk}_i\) for the encryption of shadows. \\
\textsc{distribute\_shares}(\newline\phantom{m} \(\{\overline{u_{ij}}\}_{j \neq i}, \pk_i, \{A_{i,k}\}_{k=1}^{t}\)) & After the registration phase, each member \(i\) calls this function to submit encrypted shadows for all other members and verification values for their polynomial. A hash of the arguments is stored in addition to \(\pk_i\).\\
\textsc{submit\_dispute}(\newline\phantom{m} \(\{\overline{u_{ij}}\}_{j \neq i}, \pk_i, \{A_{i,k}\}_{k=1}^{t}, k_{ij}, \pi(k_{ij})\)) & If a member \(j\) finds one of their received shadows to be invalid, they call this function and submit the \Fdistshares broadcast of the offending member \(i\) as well as the corresponding symmetric encryption key and correctness proof. The contract checks the integrity of the broadcast via its stored hash, validates the proof, decrypts the shadow, checks its validity via \cref{eq:shadow} and disqualifies the offending member if the shadow is indeed invalid. If any of these steps fails, the dispute has no effect. \\
\Fgenmpk & Once the dispute phase ends, this function must be called once by a member to generate the master public key \(\mpk\) from the contributions of the remaining, qualified members. \\
\textsc{submit\_secret}(\(\sk_i, s_i\)) & At the time of disclosure, members call this function to broadcast both \(\sk_i\) and \(s_i\). \\
\textsc{submit\_msk}(\(\msk\)) & Once enough members broadcast their share, a member can reconstruct the master secret key and submit it with this function. The contract checks the submission against the master public key determined in \Fgenmpk. \\
\bottomrule
\end{tabularx}
\end{table*}
With these optimisations and the extension of a scheduled key recovery, ETHTID proceeds in five phases: \emph{Setup}, \emph{Registration}, \emph{Share Distribution}, \emph{Dispute} (if needed), and \emph{Reconstruction}.
We first consider an execution without misbehaving parties before examining different cases of misbehaviour and the contract's ability to handle them.
\cref{tab:tidinterface} gives an overview of the functions the TID contract provides and a brief explanation of what their execution encompasses.

\subsection{Execution without Misbehaviour}
First, the \emph{initiator} deploys the ETHTID contract by submitting a corresponding transaction to the Ethereum peer-to-peer network.
As a part of the deployment transaction, the initiator also sets the threshold \(t\) as a fraction of the council members registered by the end of the next phase.
The initiator also sets the schedule for the subsequent phases at this point, most crucially the time for the coordinated recovery \(r\), via block heights of the Ethereum blockchain.
All of these settings are write-once and cannot be changed afterwards.
In practice, the initiator would also supply the contract with a reward to incentivise correct participation.

During the \emph{Registration} phase, council members call \Fregister and submit a public key to be used for the confidential exchange of shadows.
For the sake of simplicity, we use a first-come-first-served method here, but more intricate approaches could be used in practice, notably to defend against Sybil attacks.
Depending on the incentive scheme, council members would also submit a deposit as part of their registration.

With the beginning of the \emph{Share Distribution} phase, registration is no longer possible.
Each registered council member \(i\) draws a random individual secret \(\sk_i\), embeds it into their random individual polynomial \(p_i = \sk_i + \sum_{k=1}^{t} a_{i,k} X^{k}\) of degree \(t\), and generates verification values as described in \cref{sec:dkg}: \(\pk_i = \gen^{\sk_i}, \{A_{i,k} = \gen^{a_{i,k}}\}_{k \in \{1,\ldots,t\}}\).
Member \(i\) encrypts the shadow \(u_{ij} = p_i(j)\) via a one-time pad based on the symmetric encryption key \(k_{ij}\) (see \cref{eq:symkey}) using a cryptographic hash function \(\textsc{H}(k_{ij} || j)\).
This way, the shadows \(u_{ij}\) and \(u_{ji}\) are not encrypted with the same one-time pad but both recipients are still able to decrypt them.
Each member then calls \Fdistshares to broadcast both the encrypted shadows for other members \(\{\overline{u_{ij}}\}_{j \neq i}\) as well as the verification values from above.
The smart contract stores a hash over the encrypted shadows and verification values for a possible dispute.
In this way both the \((n,n)\) and \((t+1,n)\)-sharing of \(\msk\), as described in \cref{sec:dkg}, are performed in a single step.
Note, however, that council members who broadcast invalid shadows can still be disqualified via the \emph{Dispute} phase that we examine closer in the next section.

With the end of the Dispute phase, all remaining council members can combine their shadows to generate their threshold share \(s_i\) of the master secret key \(\msk\).
Council members are meant to keep both \(s_i\) and their individual secret \(\sk_i\) private until the scheduled time for disclosure and only then broadcast both via \Fsecret.
As soon as \(t+1\) members do this, anyone can compute the master secret key \(\msk\) via \cref{eq:lagrange} and submit it to the contract via \Fsubmsk to conclude the functional part of the protocol.
It is worth noting that the smart contract performs no checks on \(\sk_i\) or \(s_i\) as part of \Fsecret.
While the former is trivial, the latter is rather infeasible as it would require the resubmission of the \Fdistshares broadcast messages of \emph{all} qualified council members and the combination of \emph{all} verification values into those of the group polynomial.
We argue that neither of these checks are necessary in practice, since the council is collectively under pressure to produce the master secret key in order to earn their reward and reclaim their security deposit.

\subsection{Misbehaviour Detection and Dispute Handling}
The first general class of misbehaviour that smart contracts handle rather easily is inactivity.
If a member is expected to call a function but fails to do so within a certain amount of time, designated by block height in Ethereum, a smart contract can notice the passing of a predefined deadline during a transaction and act accordingly.
There are two opportunities for a council member to be inactive in ETHTID: during the Share Distribution and the Reconstruction phase.
If a council member registers but fails to call \Fdistshares, they would lose their security deposit and the protocol proceeds without them.
In the case that too many council members remain inactive in this way, the entire protocol would have to be restarted with a new council.
Due to the established threshold sharing of \(\msk\), up to \(n-t-1\) council members can refrain from calling \Fsecret without consequence, as the remaining members are still able to complete the reconstruction.
Consequently, if one more member remains inactive during this phase, \(\msk\) cannot be recovered.
In practice, this outcome would be discouraged via an incentive scheme \cite{yakira2019rational}.

The more interesting case of misbehaviour revolves around the validity of shadows broadcast via \Fdistshares, which our implementation inherits from ETHDKG \cite{ethdkg}.
Based on the verification values that a council member must broadcast alongside the encrypted shadows, other members can verify the correctness of all received and decrypted shadows via \cref{eq:shadow}.
If a member \(j\) detects an invalid shadow, they should call \Fdispute and resubmit the broadcast of the accused member \(i\) along with the corresponding symmetric encryption key \(k_{ij}\) and a zero-knowledge correctness proof.
The ETHTID contract verifies the integrity of the broadcast via the previously stored hash, verifies the correctness proof, decrypts the shadow in question and checks its validity via \cref{eq:shadow}.
If the shadow is indeed invalid, the accused member is disqualified from the remaining protocol and their security deposit could be destroyed or redistributed.
It is worth noting that unfounded disputes have no consequences other than the execution costs for the accusing member.
Similarly, a single valid dispute suffices to disqualify a misbehaving member, regardless of how many invalid shadows they may have sent.
In such a case, only the first dispute would involve the costly verification of a shadow, whereas any subsequent disputes against the same offending member would be recognised as moot and not incur any significant cost.

\section{Evaluation}\label{sec:eval}
With the functionality of the smart contract described in the previous section, we now examine the deployment and execution costs of the proposed construction.
We performed this evaluation with Ganache\footnote{\url{https://www.trufflesuite.com/ganache}} (v2.13.2) and compiled the smart contract with solc (v0.5.17).
At the time of writing, the active Ethereum hard fork was Muir Glacier, so we set our local development blockchain accordingly.
Recall that the Istanbul hard fork, which preceded Muir Glacier, included gas cost reductions for the precompiled contracts that we use for elliptic curve operations.
We adapted the accompanying Python application of ETHDKG \cite{ethdkg} in conjunction with its smart contract (see \cref{sec:appendix}).
The Python application serves to both check the smart contract operations for correctness as well as automate the gas cost evaluation.

Using the setup from above, we compiled and deployed our contract with thresholds of \(t=\left\lceil n/2 \right\rceil-1\) and \(t=\left\lceil 2n/3 \right\rceil-1\) before executing the protocol described in \cref{sec:ethtid} with a varying number of council members and recording the gas costs for individual transactions.
Since we observed very regular costs, we capped our evaluation at \(n=256\).

While the gas costs are constant, barring any improvements to the contract or gas price altering hard forks like Istanbul, the monetary costs are subject to market forces and network utilisation and, thus, fluctuate over time.
For a better intuition, we report execution costs in both gas and USD, using the daily average exchange rates of 1st April 2021 as reported by Etherscan\footnote{\url{https://etherscan.io}}: \SI{1967.67}[USD\ensuremath{\,}]{} per Ether and a gas price of \SI{191.87e-9}[ETH\ensuremath{\,}]{}.
It should be noted that both of these exchange rates have risen significantly since the end of 2020 and should therefore only be considered a snap shot.
With future improvements to Ethereum as part of the ETH2.0 road map, transaction costs are expected to fall again.

\begin{table}[!t]
    \centering
    \caption{Cost of functions independent of threshold \(t\) and number of council members \(n\). Conversion rates: \SI{1967.67}[USD\ensuremath{\,}]{} per Ether, \SI{191.87e-9}[ETH\ensuremath{\,}]{} per gas.}
    \begin{tabularx}{0.75\linewidth}{lRR}
    \toprule
    Function & Gas & USD\\
    \midrule
    Contract Deployment & \num{1881722} & 710.42 \\
    \Fregister & \num{106407} & 40.17 \\
    \Fsecret & \num{25196} & 9.51 \\
    \Fsubmsk & \num{52225} & 19.72 \\
    \bottomrule
    \end{tabularx}
    \label{tab:const}
\end{table}
\Cref{tab:const} shows the execution costs of functions that are independent of the threshold \(t\) and council size \(n\).
Recall that \Fsecret is merely a broadcast of two values without any checks and \Fsubmsk only verifies that the submitted master secret key is consistent with the previously generated master public key.

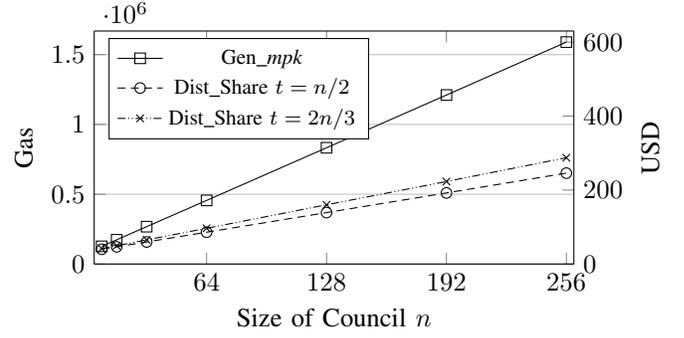
\begin{figure}[!t]
    \centering
    \begin{tikzpicture}
\pgfplotsset{set layers}
\begin{axis}[
    scale only axis,
    xmin=4,xmax=260,
    xtick={64, 128, 192, 256},
    ymin=0,ymax=1669152,
    ymajorgrids = true,
    axis y line*=left,
    xlabel=Size of Council \(n\),
    ylabel=Gas,
    legend pos=north west,
    legend style={font=\footnotesize},
    legend entries={Gen\_\textit{mpk}, Dist\_Share \(t=n/2\), Dist\_Share \(t=2n/3\)},
]
\addplot[color=black,
    mark=square]
    coordinates{(8,126763)(16,173714)(32,267663)(64,455753)(128,832701)(192,1210673)(256,1589669)}; 
\addplot[color=black,
    mark=o,densely dashed,every mark/.append style={solid}]
    coordinates {(8,105730)(16,123167)(32,158074)(64,228003)(128,368356)(192,509313)(256,651065)}; 
\addplot[color=black,
    mark=x,densely dashdotdotted,every mark/.append style={solid}]
    coordinates {(8,110788)(16,130751)(32,173217)(64,255848)(128,424323)(192,591126)(256,761382)}; 
\end{axis}
\begin{axis}[
    scale only axis,
    xmin=4,xmax=260,
    ymin=0,ymax=630.17,
    axis y line*=right,
    axis x line=none,
    ylabel=USD,
]
\end{axis}
\end{tikzpicture}
    \caption{Execution costs of \Fgenmpk (independent of \(t\)) and \Fdistshares with threshold ratios of \(t=\left\lceil n/2 \right\rceil-1\) and \(t=\left\lceil 2n/3 \right\rceil -1\). Conversion rates: \SI{1967.67}[USD\ensuremath{\,}]{} per Ether, \SI{191.87e-9}[ETH\ensuremath{\,}]{} per gas.}
    \label{fig:sharedist}
\end{figure}
\Cref{fig:sharedist} shows the execution costs for the \Fdistshares broadcast per council member as well as the execution cost of \Fgenmpk.
These costs of \Fdistshares scale in both the size of the council \(n\) and the threshold \(t\) since the broadcast consists of \(n-1\) encrypted shadows and \(t+1\) verification values.
Even though only one verification value and a hash over the broadcast payload is persistently stored on-chain, sending all this data with a transaction still incurs costs.
The costs for \Fgenmpk are independent of the threshold since it only involves combining the \(\pk_i\) of all \(n\) qualified council members.

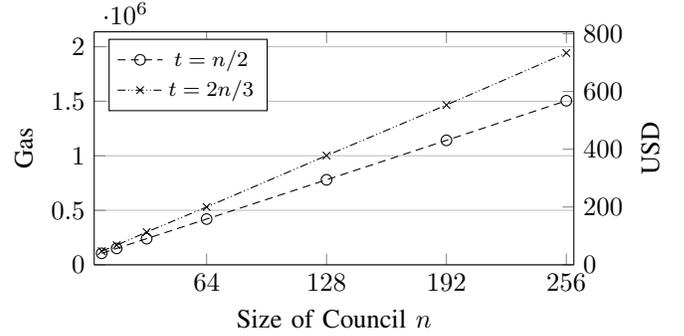
\begin{figure}[!t]
    \centering
    \begin{tikzpicture}
\pgfplotsset{set layers}
\begin{axis}[
    scale only axis,
    xmin=4,xmax=260,
    xtick={64, 128, 192, 256},
    ymin=0,ymax=2136859,
    ymajorgrids = true,
    axis y line*=left,
    xlabel=Size of Council \(n\),
    ylabel=Gas,
    legend pos=north west,
    legend style={font=\footnotesize},
    legend entries={\(t=n/2\), \(t=2n/3\)},
]
\addplot[color=black,
    mark=o,densely dashed,every mark/.append style={solid}]
    coordinates {(8,105572)(16,150361)(32,239954)(64,419464)(128,779623)(192,1141491)(256,1504658)}; 
\addplot[color=black,
    mark=x,densely dashdotdotted,every mark/.append style={solid}]
    coordinates {(8,125606)(16,180368)(32,300108)(64,530034)(128,1001441)(192,1465837)(256,1942599)}; 
\end{axis}
\begin{axis}[
    scale only axis,
    xmin=4,xmax=260,
    ymin=0,ymax=806.74,
    axis y line*=right,
    axis x line=none,
    ylabel=USD,
]
\end{axis}
\end{tikzpicture}
    \caption{Execution costs of \Fdispute with threshold ratios of \(t=\left\lceil n/2 \right\rceil-1\) and \(t=\left\lceil 2n/3 \right\rceil -1\). Conversion rates: \SI{1967.67}[USD\ensuremath{\,}]{} per Ether, \SI{191.87e-9}[ETH\ensuremath{\,}]{} per gas.}
    \label{fig:dispute}
\end{figure}
\Cref{fig:dispute} shows the execution costs for a valid \Fdispute transaction.
The brunt of these costs are caused by the evaluation of \cref{eq:shadow}, which scales with threshold \(t\) as it determines the degree of the sharing polynomials.
Since the \Fdistshares broadcast of an offending council member must be resubmitted as part of the dispute, the council size \(n\) has a very slight influence as well.
To see this, compare the costs of \(n=192, t=\left\lceil 2n/3 \right\rceil -1 = 127\) of \num{1465837} gas and \(n=256, t=\left\lceil n/2 \right\rceil-1=127\) of \num{1504658} gas.
Since the dispute mechanism is unchanged compared to ETHDKG \cite{ethdkg}, this evaluation incidentally also shows the effect of the cost reductions to the elliptic curve operations that were part of the Istanbul hard fork.

Based on these measurements, we can determine the following lower bounds for the deployment and execution costs, based on the selection of \(n\) and \(t\):
\begin{align*}\label{eq:costs}
    \textit{Gas}^{t=\left\lceil n/2 \right\rceil-1}(n) &= 2100565 + 127100n \\
    \textit{USD}^{t=\left\lceil n/2 \right\rceil-1}(n) &= 793.04 + 47.98n \\
    \textit{Gas}^{t=\left\lceil 2n/3 \right\rceil -1}(n) &= 2101773 + 131723n \\
    \textit{USD}^{t=\left\lceil 2n/3 \right\rceil -1}(n) &= 793.50 + 49.73n.
\end{align*}
Note that these bounds represent a best case scenario without any disputes and where each phase is completed with the minimally necessary transactions:
One contract deployment, \(n\) calls of \Fregister and \Fdistshares, one call of \Fgenmpk, \(t+1\) calls of \Fsecret, and one call of \Fsubmsk.

To demonstrate the amount of gas saved by our adaptations, we evaluated both ETHDKG and ETHTID on the Muir Glacier hard fork in two secnarios: A happy case where everything goes as planned and no council member misbehaves or becomes inactive, similar to the description in the previous paragraph; and a sad case where one council member distributes invalid shares and all but the minimally required \(t+1\) council members become inactive after the distribution of shares.
It is important to note that ETHDKG lacks the functionality necessary for a scheduled reconstruction of \(\msk\).
However, the added deployment costs should be minimal and our measurements of \cref{tab:const} show that the added execution costs are very low as well.
The costs for \Fsubmsk would be higher by a constant amount for ETHDKG since an additional pairing check would be necessary but it would still not present a significant increase to the overall costs.

\begin{figure}[!t]
    \centering
    \begin{tikzpicture}
\pgfplotsset{set layers}
\begin{axis}[
    scale only axis,
    xmin=4,xmax=260,
    xtick={64, 128, 192, 256},
    ymin=0,ymax=353031483, 
    ymajorgrids = true,
    axis y line*=left,
    xlabel=Size of Council \(n\),
    ylabel=Gas,
    legend pos=north west,
    legend style={font=\footnotesize},
    legend entries={ETHDKG Happy, ETHDKG Sad, ETHTID Happy, ETHTID Sad},
]
\addplot[color=black,
    mark=o,densely dashed,every mark/.append style={solid}]
    coordinates {(8,6293050)(16,10095757)(32,18539060)(64,38784289)(128,92760635)(192,164830653)(256,255132399)}; 
\addplot[color=black,
    mark=x,densely dashed,every mark/.append style={solid}]
    coordinates {(8,6524441)(16,10670687)(32,20251296)(64,44573681)(128,113939543)(192,211086618)(256,336220460)}; 
\addplot[color=black,
    mark=o]
    coordinates{(8,3858590)(16,5982413)(32,11068138)(64,24598212)(128,65148856)(192,123781676)(256,200661536)}; 
\addplot[color=black,
    mark=x]
    coordinates{(8,3964162)(16,6132774)(32,11308092)(64,25017676)(128,65928479)(192,124923167)(256,202166194)}; 

\end{axis}
\begin{axis}[
    scale only axis,
    xmin=4,xmax=260,
    ymin=0,ymax=135398.19,
    axis y line*=right,
    axis x line=none,
    ylabel=USD,
]
\end{axis}
\end{tikzpicture}
    \caption{Total execution costs of ETHDKG and ETHTID. Happy case: Everything goes as expected. Sad case: One incorrect share distribution and dispute and only \(t+1\) council members remain to complete each protocol. Conversion rates: \SI{1967.67}[USD\ensuremath{\,}]{} per Ether, \SI{191.87e-9}[ETH\ensuremath{\,}]{} per gas.}
    \label{fig:comparison}
\end{figure}
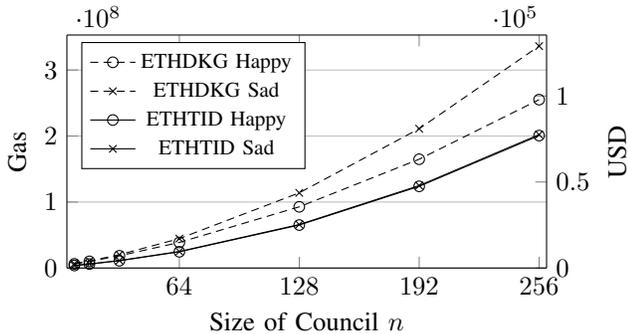
\begin{figure}[!t]
    \centering
    \begin{tikzpicture}
\pgfplotsset{set layers}
\begin{axis}[
    scale only axis,
    ybar,
    bar width=8pt,
    xtick={1,2,3,4,5,6,7},
    xticklabels={8,16,32,64,128,192,256},
    ymin=0,ymax=50,
    ymajorgrids = true,
    xlabel=Size of Council \(n\),
    ylabel=Gas Savings (\%),
    legend pos=south west,
    legend style={font=\footnotesize},
    legend entries={Happy Case, Sad Case},
]
\addplot[color=black,
    ]
    coordinates{(1,38.6848984)(2,40.7432944)(3,40.2982783)(4,36.5768649)(5,29.766699)(6,24.9037277)(7,21.3500376)}; 
\addplot[color=black,
    pattern=north east lines]
    coordinates{(1,39.2413542)(2,42.5269057)(3,44.1611441)(4,43.8734351)(5,42.1373149)(6,40.8190021)(7,39.8709424)}; 

\end{axis}
\end{tikzpicture}
    \caption{Relative cost savings achieved by ETHTID compared to ETHDKG. Happy case: Everything goes as expected. Sad case: One incorrect share distribution and dispute and only \(t+1\) council members remain to complete each protocol.}
    \label{fig:savings}
\end{figure}
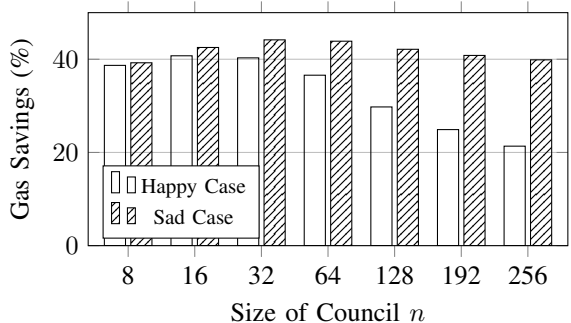
\cref{fig:comparison} shows a direct comparison of the total costs of ETHDKG and ETHTID and \cref{fig:savings} illustrates the relative gas savings.
It is very clear that the additional broadcast and optional reconstruction phases of ETHDKG for the biasing countermeasure due to Neji et al.\cite{neji2016distributed} are the main sources for gas savings, as demonstrated by the difference between happy and sad case.
Nevertheless, even in the happy case, we observe gas savings of 20--40\%.
It is also noteworthy how the happy and sad cases for ETHTID show very similar costs.
This is due to the only difference in execution being a call of \Fdispute to disqualify a member who performed an invalid share distribution.

\section{Discussion and Future Work}\label{sec:disc}
First and foremost, we can deduce from the results of \cref{sec:eval} that ETHTID is indeed deployable on Ethereum and able to execute the distributed key generation and threshold sharing by Feldman \cite{feldmanPracticalSchemeNoninteractive1987} as well as a coordinated reconstruction.
With up to \(256\) council members, all of the mandatory transactions or optional disputes stay far below the current block gas limit of \num{15000000}, although the current market situation makes practical executions financially challenging, except for small council sizes.
The results outlined above also allow estimating the expected total execution costs for any choice of council size \(n\) and certain threshold ratios \(t\), thus allowing a quantitative cost-benefit analysis for a particular instantiation.
A low threshold makes it difficult to prevent the coordinated disclosure but also lowers the bar for an attacker to obtain the master secret key through bribery or compromise of council members.
A high threshold raises this bar but also lowers the number of members that can drop out of the protocol through inaction or valid dispute before the secret key is lost.
The incentive scheme by Yakira et al.~\cite{yakira2019rational} also reflects this trade-off with the ratio \(t/n\) and could be used in conjunction with ETHTID with minor adjustments.

Currently, the ETHTID contract is only able to run the TID protocol once.
With some adjustments, it could be made reusable, which would allow an amortisation of deployment costs.
However, in order to reuse prior council member registrations, the one-time pad construction for the distribution of shadows would have to be altered, as it would currently weaken with every reuse.
A more challenging avenue for future work revolves around a \enquote{snitching} mechanism to discourage council members from colluding by sharing information that is meant to be kept secret.
However, such a mechanism could also entice council members to collude, recover the secret of a particular member, and frame them.
A similar functionality, albeit without any negative consequences for the targeted party, is a core component of ETHDKG \cite{ethdkg}.
In order for an anti-collusion mechanism like the one by Dong et al.~\cite{dong2017betrayal} to be applicable, information that would have to be shared in a collusion must be efficiently verifiable by the smart contract.

It is important to discuss both the assumptions we base our constructions on as well as the freedoms it provides in its applications.
Mainly, the selection of council members is a linchpin that can render all efforts meaningless if not done with the utmost care.
Since systems like Ethereum lack identity management that would prevent an individual from hiding behind multiple seemingly unrelated accounts, selecting a subset of \(n\) threshold council members where at most \(t\) collude becomes a challenge in itself.
Recently, proof-of-work-based Sybil defences that could be applicable to our concept have been developed \cite{gupta2019peace} that are currently being refined and improved \cite{gupta2020togcom}.
An alternative to the first-come-first-served selection method we used for simplicity would be a manual preselection of council members by the initiator.
For example, if conventional notaries or respected institutions or individuals offer their service as a temporary secret keeper for a procedure such as the one presented here, letting the initiator select council members might be a viable option.
Similarly, a form of \enquote{reputation and collateral} system that spans across multiple instantiations of the TID protocol could be used to build trust in participants over time.
Users of the temporally decoupled key pair could then judge for themselves whether or not they are satisfied with the selection of secret keepers.

While the responsibility of selecting council members seems daunting, the freedom that the protocol described here yields is also noteworthy.
By encrypting a random symmetric key with the ElGamal master public key, the submissions of users can have any size or structure but their disclosure can still be coordinated by the threshold council holding the master secret key shares.
With a reference to the coordinating smart contract attached to encrypted submissions, no further actions by users is required for the scheduled disclosure.
It is worth noting here that enforcing the correctness or even usefulness of individual submissions is not achieved by the TID contract described here.
Checking and possibly rewarding submissions as part of an actual use case is a problem that can be tackled now that the coordinated disclosure is practically feasible.

One aspect of disclosure that is fundamental to this kind of procedure is that, after \(t\) council members publish their share, all remaining \(n-t\) participants are in the advantageous position of being able to privately reconstruct the decryption key and thus learn the contents of all submissions before anyone else.
An attacker who wishes to prevent the reconstruction of the master secret key and the disclosure of submissions would have to coerce these \(n-t\) members to keep their share hidden or to destroy it.
Compared to cryptographic protocols where only one participant enjoys this privileged role and can thus prevent the protocol from completing by not cooperating, this is certainly preferable in terms of resilience.
The incentive scheme by Yakira et al.~\cite{yakira2019rational} deals with this by rewarding or punishing the entire council as a whole, depending on whether or not the reconstruction completes in time.
This is quite fortunate as determining guilt or innocence on a per-member basis is currently infeasible, due to the effort in combining all verification values to check the validity of a submitted share.
In light of this, selecting appropriate values for \(n\) and \(t\) becomes even more interesting.

On the topic of attackers, intertwining cryptography and economics in this fashion presents various opportunities to expand existing attacker models that are currently rather binary, with participants either adhering to the protocol completely or being under the full control of an attacker.
Similarly, the goal of an attacker is usually quite singular in these models, be it extracting a secret or distinguishing between two messages, to name two common examples.
With economic incentives in a practical setting, both the states of participants and the goals of an attacker become more varied.
For example, council members could follow the protocol but look to sell their secrets and shares to an attacker to maximise their profit.
Attackers could also look to cause as much disruption as possible for a given budget they are willing to lose.
Handling such scenarios without giving council members or users perverse incentives is the main challenge when adapting an incentive scheme for the protocol presented here.

Lastly, we look towards future developments in Ethereum and possible improvements to our implementation they could present.
Currently, only the Barreto-Naehrig curve~\cite{BNCurve} is supported via precompiled contracts in Ethereum.
While its bilinearity property makes it quite useful for certain applications, it would not have been the first choice for our implementation if more suitable alternatives were available.
Using other curves via custom implementations in Solidity is technically possible, but prohibitively expensive in execution.
More precompiled contracts for elliptic curve cryptography are currently being discussed\footnote{\url{https://eips.ethereum.org/EIPS/eip-1962}}, which would include curves that are more suitable to our use case like secp256k1 or Ed25519.

\section{Conclusion}\label{sec:concl}
In this paper, we presented ETHTID, an Ethereum smart contract that acts as a coordinator and arbiter of conflict for a distributed key generation, threshold sharing, and coordinated reconstruction to facilitate Threshold Information Disclosure in a context of mutual distrust.
We demonstrated the deployability of the construction experimentally and provided measurements to estimate the overall execution costs based on council size and threshold for disclosure.
With ETHTID providing a functionality that Ethereum does not offer innately, namely the coordinated disclosure of arbitrary data by mutually distrustful parties, new applications may become possible, particularly when it comes to publishing records from multiple parties independently.
While our results are generally positive, we also highlight areas where both the tools available in Ethereum as well as their application can be improved.
Simpler, more efficient, and cheaper elliptic curves would fit the presented application better.
Augmenting the presented smart contract to be reusable and refining the cryptographic construction to support a \enquote{snitching} mechanism in conjunction with an incentive scheme in order to discourage premature sharing of secrets appear to be interesting future work.

\bibliographystyle{IEEEtran}
\bibliography{IEEEabrv,references}

\appendix\label{sec:appendix}

In this appendix, we briefly explain the differences between the ETHDKG and ETHTID smart contracts.

ETHTID does not require the cryptographic constants for the \enquote{generator switch} from \(\gen\) to \(h\), so the constants \texttt{H1x}, \texttt{H1y}, \texttt{H2xi}, \texttt{H2x}, \texttt{H2yi}, \texttt{H2y} are removed.
Similarly, the events \texttt{KeyShareSubmission} and \texttt{KeyShareRecovery} are no longer needed and can be removed.
A new event \texttt{SecretSubmission} is added to simplify the reconstruction phase.
The mapping \texttt{key\_shares} is not needed anymore since the mapping \texttt{commitments\_1st\_coefficient} contains the corresponding values for the generator \(\gen\).
Next to \texttt{master\_public\_key}, EHTTID is also intended to store \texttt{master\_secret\_key}.
For the scheduling, the variable \texttt{T\_KEY\_SHARE\_SUBMISSION\_END} is no longer needed but two new variables are introduced to govern the reconstruction phase: \texttt{T\_SKEY\_RECON\_BEGIN}, where the block height for the beginning of the reconstruction phase will be stored once it is determined through the publication of \(\mpk\) and  \texttt{DELTA\_HOLD}, a hard coded constant to determine how many blocks should pass between the publication of \(\mpk\) and the beginning of the reconstruction phase. The \texttt{constructor()} is altered accordingly.

The functions \texttt{register()}, \texttt{distribute\_shares()}, and \texttt{dispute()} are unchanged whereas the functions \texttt{submit\_key\_share()} and \texttt{recover\_key\_shares()} are removed entirely. The function \texttt{submit\_master\_public\_key()} no longer needs to receive \(\mpk\) as an argument, instead the contract can construct \(\mpk\) from the mapping \texttt{commitments\_1st\_coefficient}. Additionally, the beginning of the reconstruction phase is set as \texttt{T\_SKEY\_RECON\_BEGIN = block.number + DELTA\_HOLD}. As part of this simplification, the pairing check can also be dropped. Lastly, ETHTID includes the broadcast helper function \texttt{submit\_secret()} to submit individual secrets and shares and the function \texttt{submit\_master\_secret\_key()} to persist \(\msk\) after checking it against \(\mpk\).
\end{document}